\documentclass[aps, prd,twocolumn,nofootinbib,tightenlines,floatfix,showpacs ,amssymb,superscriptaddress]{revtex4-2}
\usepackage{amsmath,graphicx}
\usepackage{mathrsfs}

\begin{document}


\title{Earth-mass primordial black hole mergers as sources for non-repeating FRBs}
\author{Can-Min Deng}
\affiliation{GXU-NAOC Center for Astrophysics and Space Sciences, Department of Physics, Guangxi University, Nanning 530004, China}
\affiliation{CAS Key Laboratory for Research in Galaxies and Cosmology, Department of Astronomy, University of Science and Technology of China, Hefei 230026, Anhui, China}


\begin{abstract}
Fast radio bursts (FRBs) are mysterious astronomical radio transients with extremely short intrinsic duration. Until now, the physical origins of them still remain elusive especially for the non-repeating FRBs. Strongly inspired by recent progress on possible evidence of Earth-mass primordial black holes,  we revisit the model of Earth-mass primordial black holes mergers as sources for non-repeating FRBs. Under the null hypothesis that the observed non-repeating FRBs are originated from the mergers of Earth-mass primordial black holes, we analyzed four independent samples of non-repeating FRBs to study the model parameters i.e. the typical charge value $q_{\rm{c}}$ and  the power index $\alpha$ of the charge distribution function of the primordial black hole population $\phi(q) \propto (q/q_{\rm{c}})^{-\alpha}$ which describe how the charge was distributed in the population.  $q$ is the charge of
 the hole in the unit of $\sqrt{G} M$, where $M$ is the mass of the hole. It turns out that this model can explain the observed data well. {Assuming the monochromatic mass spectrum for primordial black holes}, we get the average value of typical charge $\bar{q}_{\rm{c}}/10^{-5}=1.59^{+0.08}_{-0.18}$ and the power index $\bar{\alpha}=4.53^{+0.21}_{-0.14}$ by combining the fitting results given by four non-repeating FRB samples. The  event rate of the non-repeating FRBs can be explained in the context of this model, if the abundance of the primordial black hole populations with charge $q \gtrsim 10^{-6}$ is larger than $10^{-5}$  which is far below the upper limit given by current observations for the abundance of  Earth-mass primordial black holes. In the future, simultaneous detection of FRBs and high frequency gravitational waves produced by mergers of Earth-mass primordial black holes may directly confirm or deny this model.
\end{abstract}


\maketitle

\section{Introduction}
Fast radio bursts (FRBs) are mysterious astronomical radio transients with extremely short intrinsic duration $\lesssim $ ms \citep{Lorimer:2007qn,Cordes:2019cmq,Petroff:2019tty}. 
There have been more than 100 published FRBs, and most of them  are one-off events, only 20s sources have been found to repeat \citep{Petroff:2016tcr}. 
Observations of their host galaxy revealed that their surrounding environments are different \citep{Macquart:2020lln,Li:2020esc}.
A question arises, are the one-off bursts and the repeating bursts essentially the same in physical origin?  This question remains a mystery. However, some studies seem to be inclined to suggest that FRBs should  be divided into two categories intrinsically \citep{2018ApJ...854L..12P,2021ApJ...906L...5A,2021MNRAS.500.3275C}, namely repeating FRBs and non-repeating FRBs.

Thanks to abundant observation data, the study on the physical origin of the repeating FRBs has come a long way.
It is believed that the repeating FRBs are originated  from the magnetar activity \citep{2013arXiv1307.4924P,2014MNRAS.442L...9L,2014ApJ...797...70K,2016ApJ...826..226K,2017ApJ...841...14M,2017ApJ...843L..26B,2017MNRAS.468.2726K,2019MNRAS.485.4091M,2019ApJ...879....4W,2020ApJ...896..142B,2020MNRAS.498.1397L,2021FrPhy..1624503L} or the interaction of binary stars containing at least one compact object \citep{2015ApJ...809...24G,2016ApJ...829...27D,2016ApJ...823L..28G,2020MNRAS.497.1543G,2020ApJ...895L...1D,2020ApJ...897L..40D,2020ApJ...898L..55G,2020ApJ...897L..40D,2020ApJ...893L..26I,2020ApJ...890L..24Z,2021A&A...645A.122D,2021arXiv210204264K,2021arXiv210206796D}.
However, the origin of the non-repeating FRBs remains elusive. It is thought that the non-repeating FRBs are likely to originate from the catastrophic processes, such as the collapse of the neutron stars into black holes \citep{2014A&A...562A.137F,2014ApJ...780L..21Z,2016MNRAS.459L..41P} or the merger of two compact stars \citep{2013PASJ...65L..12T,2013ApJ...776L..39K,2015ApJ...814L..20M,2016ApJ...827L..31Z,2016ApJ...822L...7W,2016ApJ...826...82L,2018RAA....18...61L,2018Ap&SS.363..242L}.
Since none of the above models can perfectly explain the non-repeating FRBs, there are also some novel ideas, such as the oscillation/decay of superconducting cosmic strings \citep{2008PhRvL.101n1301V,2012PhRvD..85b3530C,2012PhRvD..86d3521C,2014JCAP...11..040Y,2015AASP....5...43Z,2017arXiv170702397B,2017EPJC...77..720Y,2018PhRvD..97b3022C,2020EPJC...80..500I}, and our proposed model of mergers of Earth-mass primordial black holes \citep{2018PhRvD..98l3016D}. As shown in \cite{2018PhRvD..98l3016D},  the Earth-mass primordial black hole mergers can be the sources for non-repeating FRBs, and the model can in principle explain all the key observational features, especially for the high event rate $\sim 10^{4} ~\rm{Gpc}^{-1} ~\rm{yr}^{-1}$ \citep{2019JHEAp..23....1D}.

Primordial black holes are thought to be produced from density fluctuations in the early universe \citep{1967SvA....10..602Z,1971MNRAS.152...75H,1974MNRAS.168..399C,1975ApJ...201....1C,1996PhRvD..54.6040G,2016JCAP...11..029Q,2017PDU....18...47G,2017PhLB..769..561C,2017JCAP...07..048D,2017PhRvD..96f3507C,2017JCAP...09..020K,2018PhRvD..97b3501B,2018JCAP...07..032B,2018JCAP...03..016F,2018PhRvD..97h3509H,2018CQGra..35w5017K,2018JCAP...07..005O,2018PhRvL.121h1306C,2010RAA....10..495K,2018CQGra..35f3001S}.
Interestingly, recent microlensing observations by the Optical Gravitational Lensing Experiment
(OGLE)  found abnormal signal that might be  evidence of existence of the Earth-mass primordial  black holes \citep{2017Natur.548..183M}. Assuming that the anomalous signal corresponds to the Earth-mass primordial black holes, then they should be about 1\% as abundant as dark matter \citep{2019PhRvD..99h3503N}. 
More interestingly, recently North American Nanohertz Observatory for Gravitational Waves (NANOGrav) claimed that they have detected a gravitational wave background signal from the NANOGrav 12.5-yr data set \citep{2020ApJ...905L..34A}.
And this gravitational wave background signal may hint the formation of planetary-mass primordial black holes \citep{2020arXiv201003976D}.
Moreover, gravitational anomalies in our Solar System seem to require the existence of a new ninth planet with mass $\sim 5-15 M_{\bigoplus}$, but the search for a ninth planet has been fruitless \citep{2016AJ....151...22B,2019PhR...805....1B}. 
Therefore, \cite{2020PhRvL.125e1103S} argues that it is entirely possible for Planet 9 to be an Earth-mass primordial black hole.

Those new results for primordial black holes mentioned above strongly encourage us to continue the research on our previously proposed model of the merger of Earth-mass primordial black holes for non-repeating FRBs \citep{2018PhRvD..98l3016D}. 
Since the non-repeating FRBs are just a flash of radio signal and have no detectable afterglow, the information we can get is very limited. Therefore, the study of their population properties is an effective way to obtain the information about their physical origin.
Thanks to the observation of FRBs by many radio telescopes in the world, several effective samples of FRBs have been accumulated.
Therefore, in this work, we plan to use the samples of those telescopes to perform an updated study to the parameters of the  model.

%
\section{Brief Review of the Model}
Following Ref. \cite{2018PhRvD..98l3016D}, the  intrinsic  time scale of the radio bursts from the mergers of primordial black holes is 
$\tau \sim 6GM/{c^3}$, where $M$ is the total mass of the binary, $G$ is the gravitational constant, $c$ is the speed of light.
This is also the time scale of the final plunge of the binary black holes after reaching the last stable orbit (LSO) $a_{\rm{LSO}} \sim 6GM/{c^2}$, where $a$ is the separation of the binary. Then the typical intrinsic frequency of the radiation can be given by
\begin{equation}
\nu \sim \frac{1}{\tau} \sim \left(\frac{M}{10^{-5} M_{\odot}}\right)^{-1} \rm{G H z}.
\label{eq:1}
\end{equation}
As one can see, the Earth-mass system corresponds to a frequency $\sim$ GHz.
For the observation duration of the bursts $W_{\rm{o b s}}$, it is the sum of intrinsic duration $\tau$, scattering $\Delta t_{\rm{s}}$, dispersion smearing $\Delta t_{\rm{D M}}$ and sampling time $\Delta t_{\rm{s a m p}}$ of the telescope \citep{Cordes:2019cmq,Petroff:2019tty}, 
\begin{equation}
W_{\rm{o b s}}=\sqrt{\tau^{2}+\Delta t_{\rm{s}}^{2}+\Delta t_{\rm{D M}}^{2}+\Delta t_{\rm{s a m p}}^{2}}.
\label{eq:2}
\end{equation}
Thus, the intrinsic duration $\tau \sim (M/10^{-5}M_{\odot}) $ ns is usually negligible in our scenario.
The energy of the bursts is given  by \citep{2018PhRvD..98l3016D}
\begin{equation}
\begin{aligned}
E  &\sim (1+\mu)^{-1} q^2Mc^2 \\
    &\simeq 10^{38} ~(1+\mu)^{-1} \left(\frac{q}{10^{-6} }\right)^{2}\left(\frac{M}{10^{-5} M_{\odot}}\right) \rm{erg},
\end{aligned}
\label{eq:3}
\end{equation}
{where $\mu $ ($ \geq 1$) is the mass ratio of the binary}\footnote{{Here is why we consider $\mu$=1 in this work. As one can see in Eq.(3), $\mu$ and $q$ are degenerate, and FRB observations alone cannot remove the degeneracy. In the future, if there are simultaneous gravitational wave observations for Earth-mass primordial black holes mergers, then  $\mu$ can be determined and degeneracy can be broken. Only then we should consider the case of  $\mu \neq 1$. Moreover, according to Eq.(3), the larger $\mu$ is for a fixed $q$, the smaller the electromagnetic energy generated by the merger. For the case where $\mu \gg 1$, the electromagnetic energy would be far below the detection limits. Finally, the mass spectrum of the primordial black holes is unknown, while the monochromatic  mass spectrum is the simplest and the most commonly assumption in the literatures.}}, 
$q$ is the charge of the black holes in unit of $\sqrt{G}M$ as defined in \cite{2018PhRvD..98l3016D}.  In this work, again, we are only considering the case of monochromatic mass spectrum for primordial black holes, namly $\mu=1$.  Combining equation (\ref{eq:1}) with equation (\ref{eq:3}), and one gets the intrinsic luminosity of the bursts 
\begin{equation}
L \sim (1+\mu)^{-1} \frac{c^{5}q^2}{6 G}  \simeq 10^{47} \left(\frac{q}{10^{-6} }\right)^{2} \rm{erg/s}.
\label{eq:4}
\end{equation}
However, as shown in equation (\ref{eq:2}),  it is difficult to know the intrinsic luminosity of a burst observationally because the duration of the burst is greatly broadened by scattering and instrumental effects.
In contrast,  the energy of the burst will not be changed by theoe effects, so we consider the energy of the burst in the practical treatment instead of luminosity.

In order to connect  with observations, we also need to calculate the event rates of the bursts. As in Ref. \cite{2018PhRvD..98l3016D},  We assume that the whole population of the primordial black holes satisfies a charge distribution which could be a function of the charge parameter itself $\phi (q) \propto dN/dq \propto (q/q_{\rm{c}})^{-\alpha}$ ,  and $q$ is naturally cut off by the PBH mass due to the requirement of theoretical consistency i.e. $q \leqslant 1$.  Therefore, $\phi (q)$ is normalized by $\int_{0}^{1} \phi(q) d q=1$. In the context of this work, the null hypothesis is that all non-repeating FRBs are originated from the mergers of the primordial black holes. The  charge distribution of the population of primordial black holes would shape the distribution of the FRBs energy and thus affect the amount of FRBs observed in a given  survey observation. 
Therefore, the most critical parameters,  in this model,  are $q_{\rm{c}}$ and $\alpha$ which describe how the charge was distributed in the primordial black hole population. And this is exactly the research objective of this work.

Following Ref. \cite{2018PhRvD..98l3016D}, for a radio telescope survey with fluence limited sensitivity $F_{th}$, field of view $\Omega$, and operational time $T$, the observational number of bursts within the range ($q, q+dq$) can be calculated as
\begin{equation} 
dN{\rm{ = }}\frac{{\Omega T}}{{4\pi }}\phi (q)dq\int_0^{{z_{\max }}(q)} {\frac{{n(z)}}{{1 + z}}} \frac{{dV(z)}}{{dz}}dz ~,
\label{eq:6}
\end{equation}
where $dV(z)$ is the comoving volume element.  And $z_{\rm max}(q)$ is the maximum redshift where the FRBs arisen from the PBH binary coalescence with $q$ can be detected by a radio telescope with fluence limited sensitivity $F_{\rm{th}}$. It is defined as \citep{2018PhRvD..98l3016D}
\begin{equation}
{F_{\rm{th}}}  = \frac{{{(1+z_{\max })q^2}{M}{c^2}}}{{4\pi D_L^2({z_{\max }})}} ~,
\label{eq:Fth}
\end{equation}
where $D_L$ is the luminosity distance. The cosmic merger rate density  of the primordial black holes binaries is
\begin{equation}
n(z)=n_{0}(1+z)^{3}\left(t_{0} / t_{\mathrm{z}}\right)^{34 / 37},
\end{equation}
where $t_{0}$ and $t_{\rm z}=\int_{z}^{\infty} [(1+z')H(z')]^{-1} dz'$ are the age of the universe at present and redshift $z$, respectively. And $n_{0}$ is the local merger rate as follows \citep{2018PhRvD..98l3016D}
\begin{equation} 
n_{0} \simeq  10^{-3} \frac{\rho}{0_{\odot}t_{0}} m^{-32/37} f^{2} (f^{2}+\sigma_{ \emph eq}^{2})^{-21/74} ~,
\label{eq-n0}
\end{equation}
where $\rho_{0}$ is the matter density of the universe at $t_{0}$, $f$ is the fraction of primordial black holes against the matter sector of the universe, $m=M/M_{\odot}$, $\sigma_{eq}=5 \times 10^{-3}$ adopted.
According to equation (\ref{eq:6})-(\ref{eq-n0}), then we can infer the absolute cumulative distribution of $q$ from a survey as 
\begin{equation}
{n_{0}(> {q_i})} = \int_{q_{i}}^{q_{m x}} n_{0} \phi(q) d q \approx \frac{{4\pi }}{{\Omega T}}\sum\limits_{{q_i}}^{{q_{\max }}} {\frac{1}{{f({q_i})}}} ~,
\label{eq:n0q}
\end{equation}
where ${q_{\max }}$ is the maximum value of $q$ in a FRBs sample, and
\begin{equation}
f(q) = {\int_0^{{z_{\max }}(q)} {{{(1 + z)}^2}\left( {\frac{{{t_0}}}{{{t_{\rm{z}}}}}} \right)} ^{34/37}}\frac{{dV(z)}}{{dz}}dz.
\end{equation}

\section{Sample selection}
Up to date, there are currently several available subsamples in the FRB catalog \citep{Petroff:2016tcr}, including Parkes, ASKAP, CHIME, and UTMOST, which contain 30, 33, 13, and 12 non-repeating FRBs, respectively.
They can be found in the FRB catalogue\footnote{See the online catalogue for FRBs: http://www.frbcat.org/} \citep{Petroff:2016tcr}. 
For FRBs without spectroscopy redshift observations, one can inferred the redshifts from their  dispersion mesures (DMs).
In the context of our mode, the host galaxy contribution to the DM is expected to be small. Therefore, we ignore the DM contribution from the host galaxy, the observed DM of an FRB can be consisted of 
\begin{equation}
\rm DM_{obs}=DM_{MW}+DM_{IGM}\;,
\end{equation}
where $\rm DM_{MW}$ is the DM distribution from the Milky Way.
The IGM portion of DM is related to the redshift of the source through
\citep{2019JHEAp..23....1D}
\begin{equation}
{\rm DM_{IGM}}(z)=\frac{3cH_0\Omega_b f_{\rm IGM} f_{e}}{8\pi G m_{\rm{p}}}\int_{0}^{z}\frac{1+z'}{E(z')}dz'\;,
\end{equation}
where $\Omega_b =0.0486$ is baryon density, $H_{0}=67.74$ is the Hubble constant, $E(z)=H(z) / H_{0}$ is the dimensionless expansion function, $m_{\rm{p}}$ is the mass of proton, $f_{\rm IGM}\sim0.83$ is the fraction of baryons in the IGM,
and $f_{e}\sim7/8$ is the free electron number per baryon in the universe \citep{2014ApJ...783L..35D}.
After deducting the DM contribution of the Milky Way, we can estimate the redshift $z$ based on the above equation.

Following Ref. \cite{2019JHEAp..23....1D}, we calculate the isotropic energy
of an FRB within the rest-frame bandwidth $({\nu _{1 }},{\nu _{2}})$ by
\begin{equation}
E_{\rm{obs}} \simeq \frac{{4\pi D_L^2}}{{1 + z}}{F_{\nu}}\int_{{\nu _1}/(1 + z)}^{{\nu _2}/(1 + z)} {{{\left( {\frac{\nu }{{{\nu _{\rm{c}}}}}} \right)}^{ - \alpha }}d\nu }  ,
\label{eq:Eiso}
\end{equation}
where ${{F_\nu }}$ is the observed fluence density, ${{D_L}}$ is luminosity distance, $(\nu/ \nu _{\rm{c}})^{ - \alpha }$ is the spectrum of FRBs, $\nu_{c}=1.352$ GHz, $\nu_{c}=1.297$ GHz,  $\nu_{c}=0.843$ GHz, and $\nu_{c}=0.6$ GHz are the typical central frequency of Parkes, ASKAP, UTMOST and CHIME, respectively. We adopt $\alpha  = 1.6$ as in \cite{2019JHEAp..23....1D}. We take ${\nu _1} = 0.4\,{\rm{GHz}}$ and ${\nu _2} = 1.8\,{\rm{GHz}}$ to uniformly correct the energy of bursts from each telescope to the uniform band.

\section{Results}
\begin{figure}[tpb]
	\centering
	\includegraphics[width = 1\linewidth]{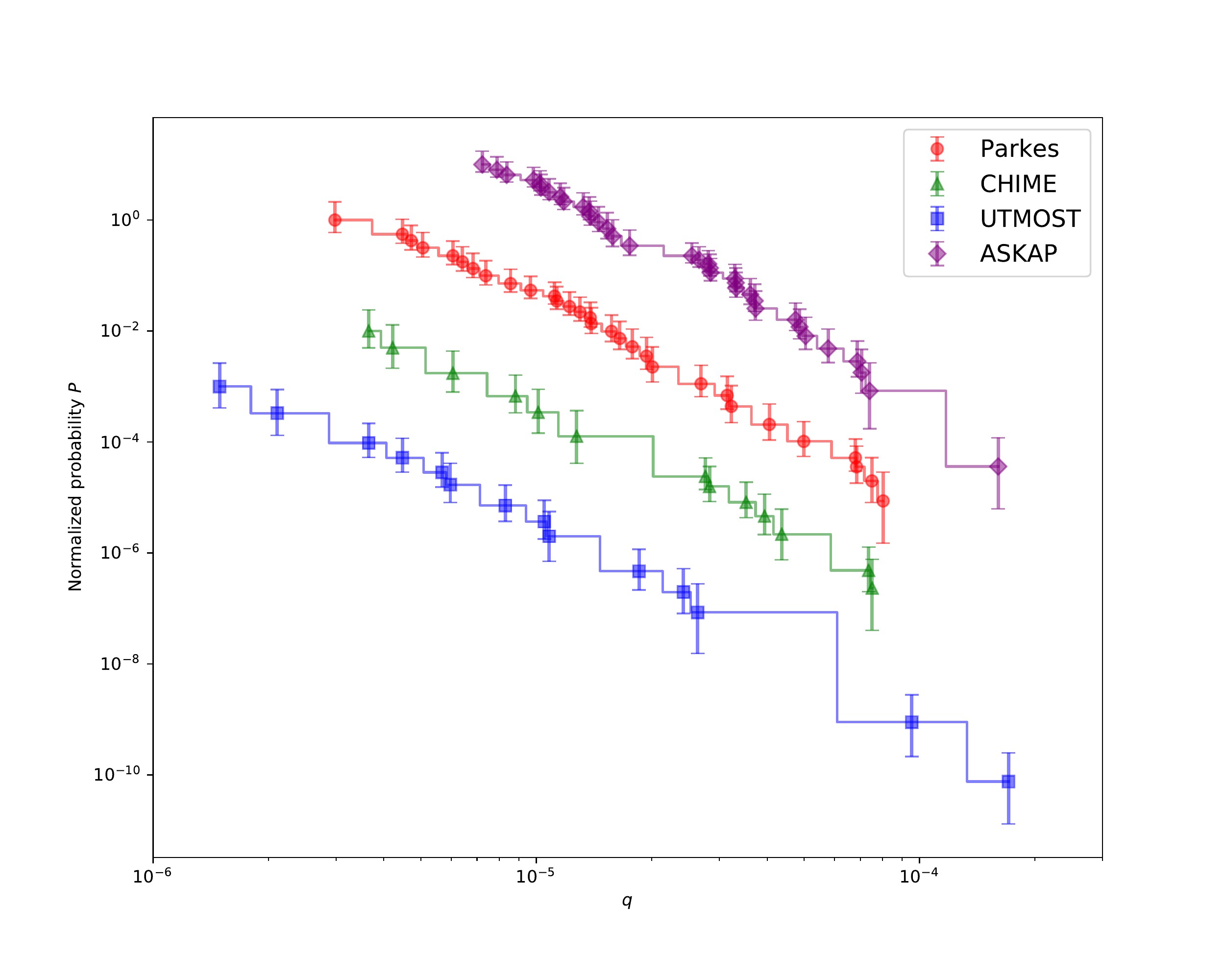}
	\caption{The normalized  cumulative distribution of $q$ derived by using formula (\ref{eq:n0q}) for the samples of CHIME (purple diamond), Parkes (red circle), ASKAP (green triangle), and  UTMOST (blue square), respectively.}
	\label{fig:1}
\end{figure}

For the samples of each telescope, formula (\ref{eq:n0q}) was applied respectively to obtain the normalized  cumulative distribution of $q$ as shown in Fig.\ref{fig:1}.
As one can see, the distribution of $q$ spans three orders of magnitude, and the distributions are very similar across different samples of telescopes.
One thing to point out is that the $F_{\rm{th}}$ required in equation (\ref{eq:Fth}) is complicated and is given by \citep{2019JHEAp..23....1D}
\begin{equation}
{F_{{\rm{th}}}} = \frac{{(\textrm{S/N}) ~\nu_{\rm{c}} {T_{{\rm{sys}}}}}}{{\beta G\sqrt {B{N_{\rm{p}}}} }}\sqrt {{W_{{\rm{obs}}}}}   ~.
\label{eq:Futh}
\end{equation}
where $\beta$ is the digitisation factor, $G$ is the system gain in ${\rm{K}}{\kern 1pt} {\kern 1pt} {\rm{J}}{{\rm{y}}^{ - 1}}$, $B$ is the bandwidth in Hz, ${{N_{\rm{p}}}}$ is the number of polarizations, ${{T_{{\rm{sys}}}}}$ is the system temperature in ${\rm{K}}$. The signal is claimed as a reliable FRB detection when the  signal to noise ratio $S/N$ reaches over some critical values,typically 8 to 10. One can see that the detection threshold is depends on the pulse width ${{W_{{\rm{obs}}}}}$.
Thus, it is difficult to define a strict observation threshold ${F_{{\rm{th}}}} $ for a given telescope due to the different duration for observed FRBs. Therefore, in this work, the ${F_{{\rm{th}}}} $ are taken directly to be the smallest value in the corresponding samples for an approximation.
And this has no adverse effect on the final results, because the final results are not sensitive to $F_{\rm{th}}$.

Furthermore, we fit the cumulative distribution of $q$ of each sample  respectively.
Accordingly, $\alpha$ is the power index and $q_{\rm{c}}$ is the typical charge of the differential distribution function $\phi(q)$, respectively.
{We adopt the fitting function as $\varPhi (q) = (q/q_{\rm{c}})^{-(\alpha-1)}$ which is the integral of $\phi(q)$.
There are two considerations to use the power-law charge distribution model.
Firstly, it can be observed from Fig. \ref{fig:1} that the distribution of $q$ apparently indicates a power-law model. Secondly, such a power-law function is the simplest model and has been widely used  in astrophysics.
}

\begin{figure}[tpb]
	\centering
	\includegraphics[width = 0.6\linewidth]{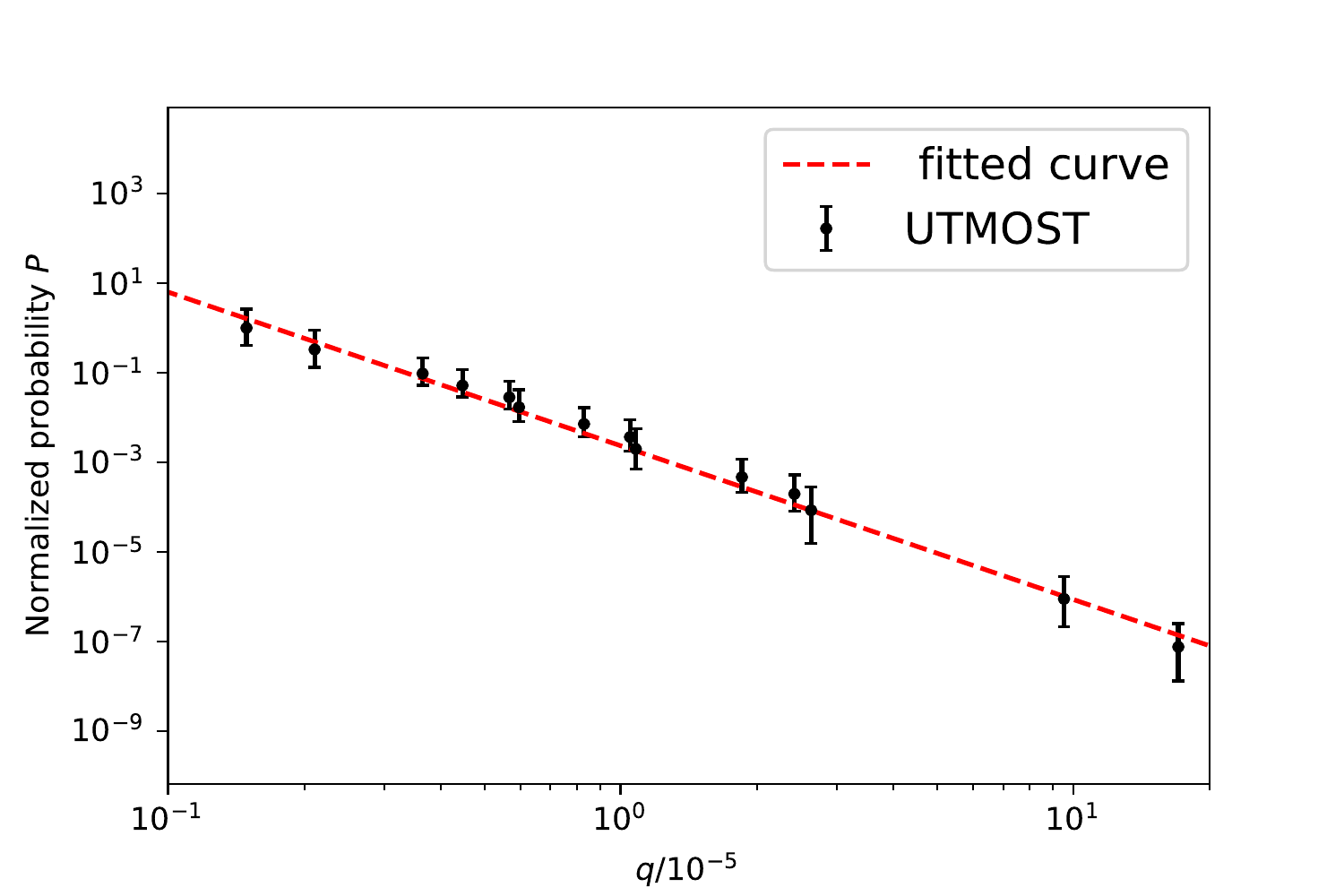}
	\includegraphics[width = 0.6\linewidth]{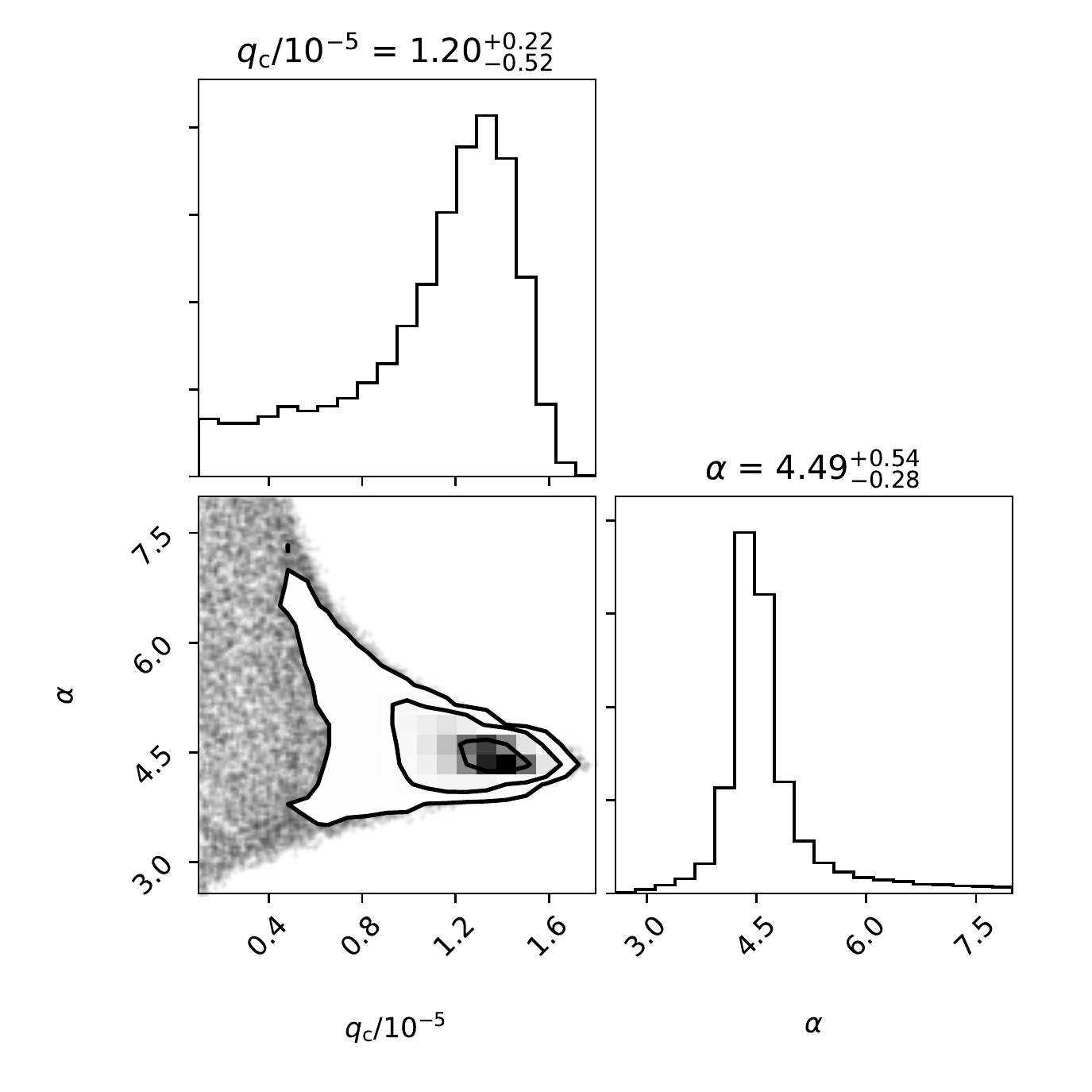}
	\caption{Upper panel: The best fit to the data. The data points are from the UTMOST sample, and the error bars are the poisson error. The red dash line denotes the best fitting curve. Lower panel: The  posterior probability distribution of the fitting parameters.}
	\label{fig:2}
\end{figure}

The fitting results for UTMOST samp  are shown in Fig. \ref{fig:2}, where the upper panel shows the best fit to the data by applying the maximum likelihood method and the lower panel shows the  posterior probability distribution of the fitting parameters obtained by using the MCMC method.
The likelihood $\mathscr{L}$ for the fit and the MCMC method is determined by 
\begin{equation}
\ln \mathscr{L} (P \mid q_{\rm{c}}, \alpha)=-\frac{1}{2} \sum_{i}\left[\frac{\left[P_{i}-\varPhi (q_i) \right]^{2}}{\sigma_{i}^{2}}+\ln \left(2 \pi \sigma_{i}^{2}\right)\right],
\end{equation}
where $P$ is the cumulative probability of $q$ and is normalized to $1$ at $q_{\rm{min}}$,  $\sigma_{i}$ is  the error of $P_{i}$. 
\begin{figure}[tpb]
	\centering
	\includegraphics[width = 0.6\linewidth]{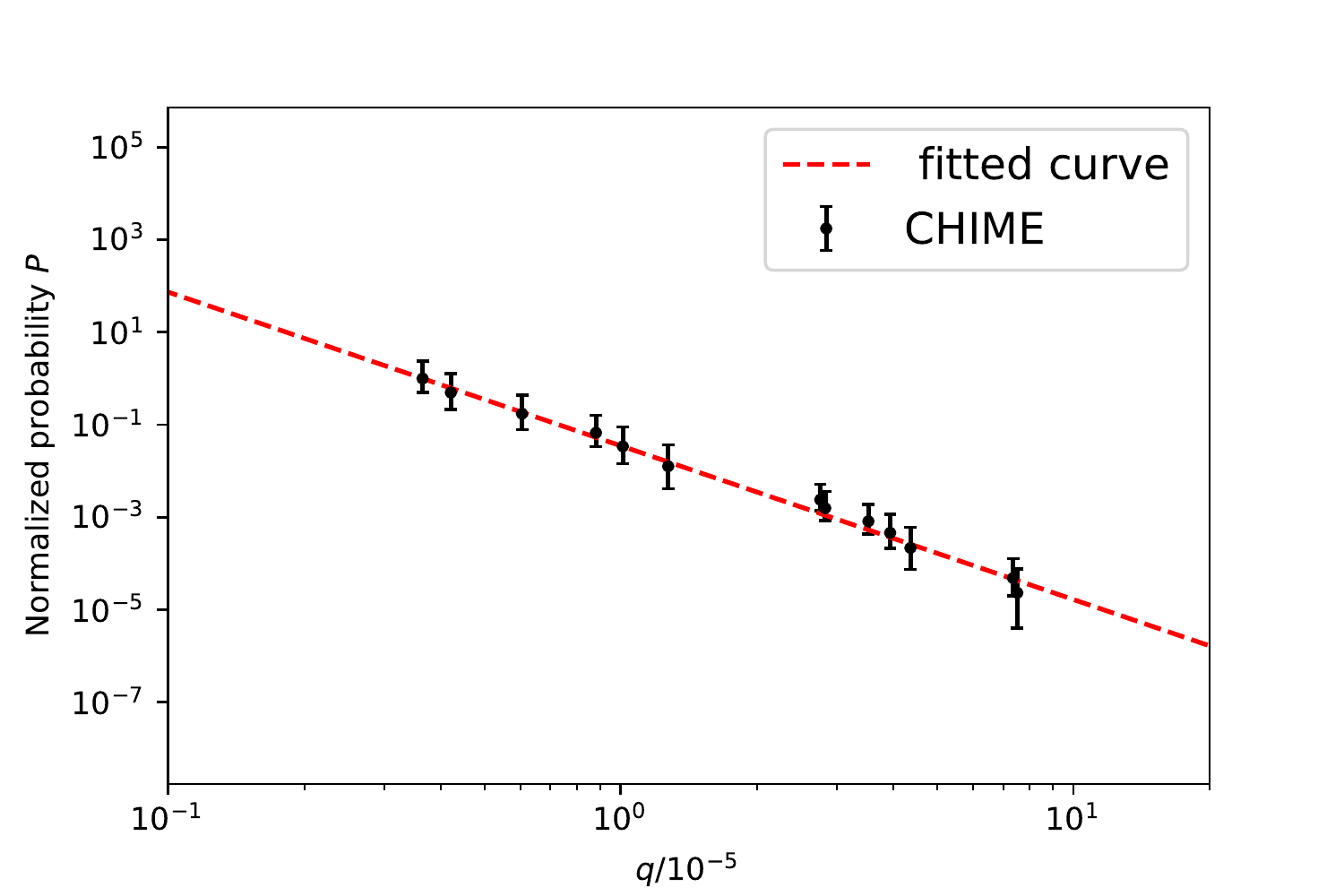}
	\includegraphics[width = 0.6\linewidth]{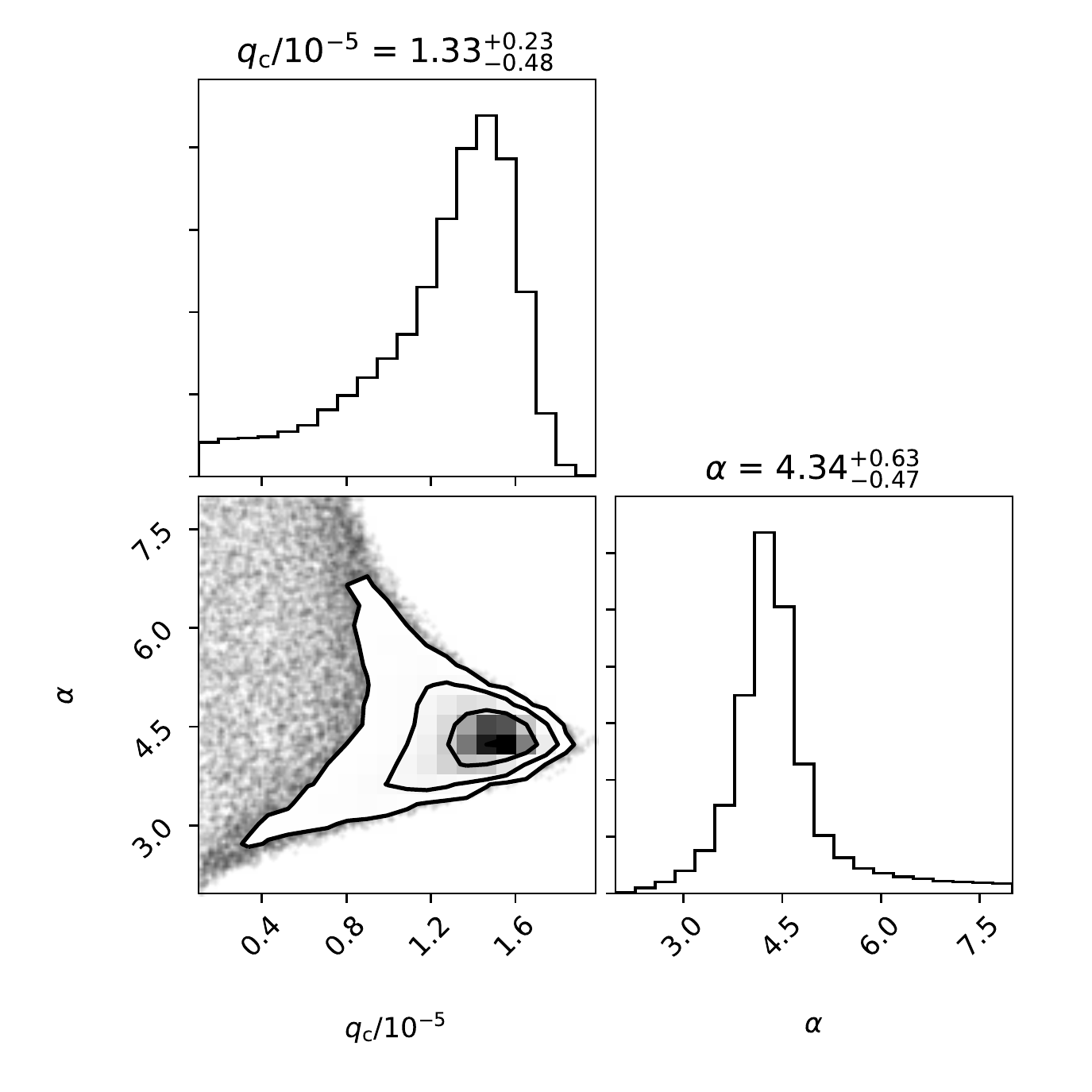}
	\caption{Upper panel: The best fit to the data. The data points are from the CHIME sample, and the error bars are the poisson error. The red dash line denotes the best fitting curve. Lower panel: The  posterior probability distribution of the fitting parameters.}
	\label{fig:3}
\end{figure}
\begin{figure}[tpb]
	\centering
	\includegraphics[width = 0.6\linewidth]{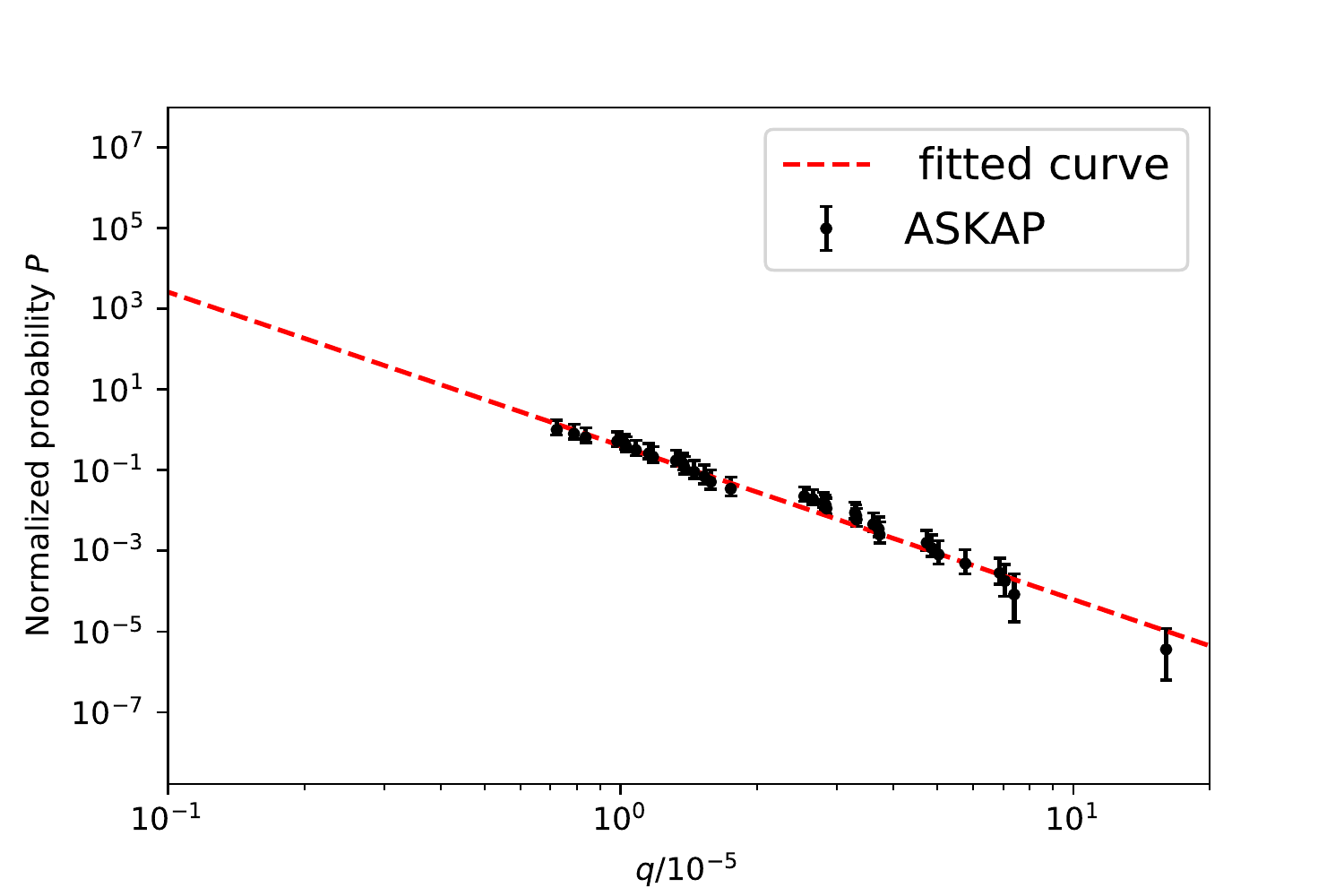}
	\includegraphics[width = 0.6\linewidth]{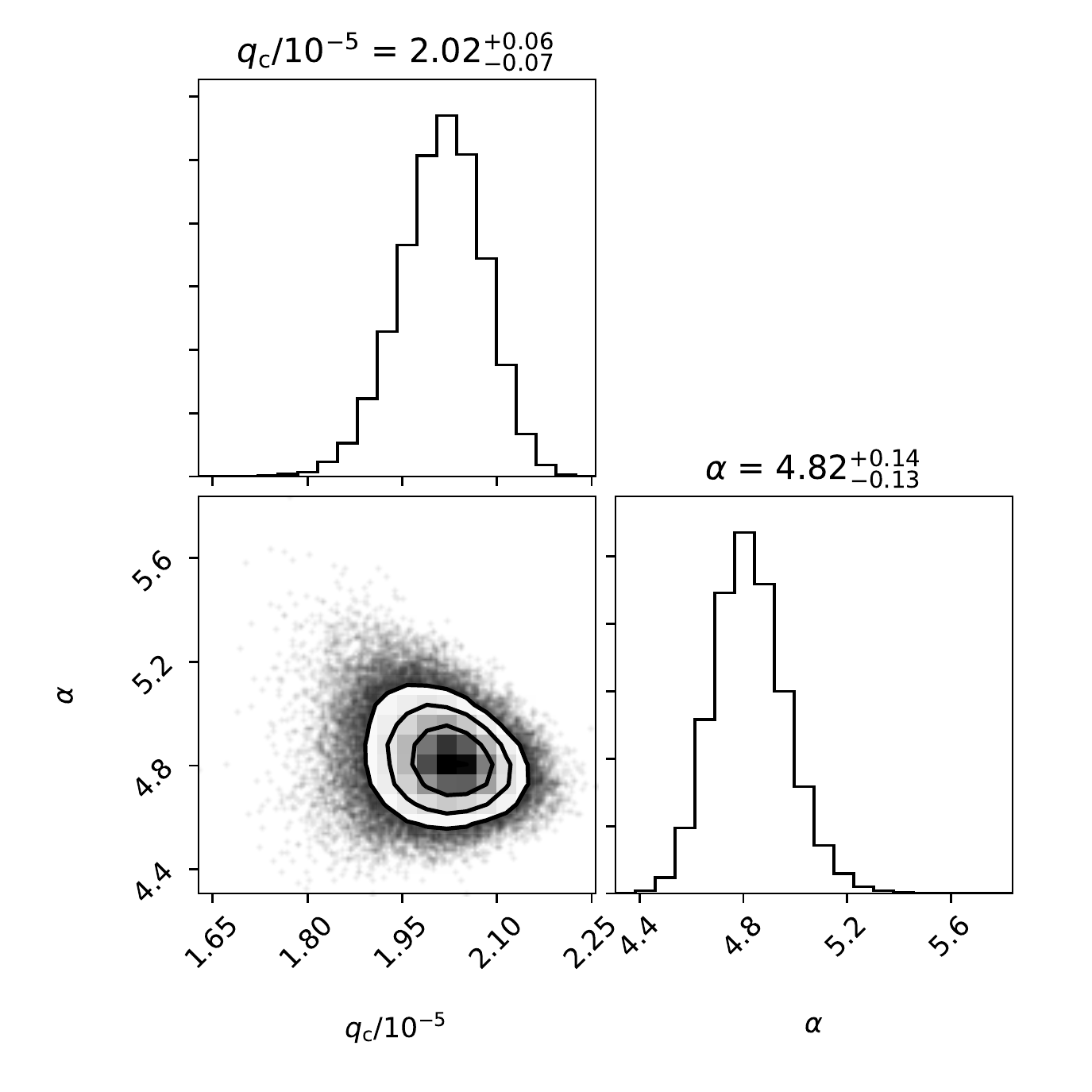}
	\caption{Upper panel: The best fit to the data. The data points are from the ASKAP sample, and the error bars are the poisson error. The red dash line denotes the best fitting curve. Lower panel: The  posterior  probability distribution of the fitting parameters.}
	\label{fig:4}
\end{figure}
\begin{figure}[tpb]
	\centering
	\includegraphics[width = 0.6\linewidth]{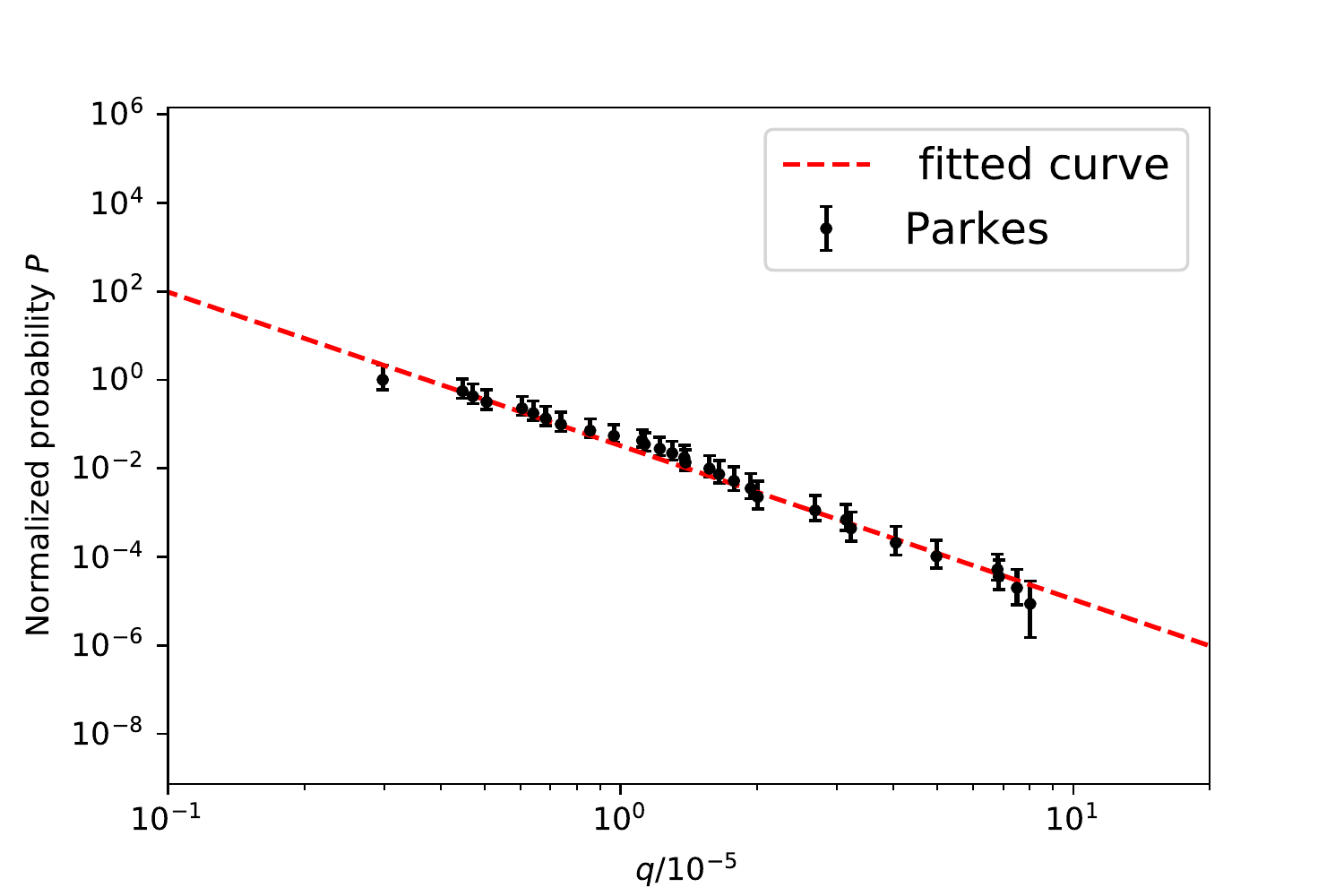}
	\includegraphics[width = 0.6\linewidth]{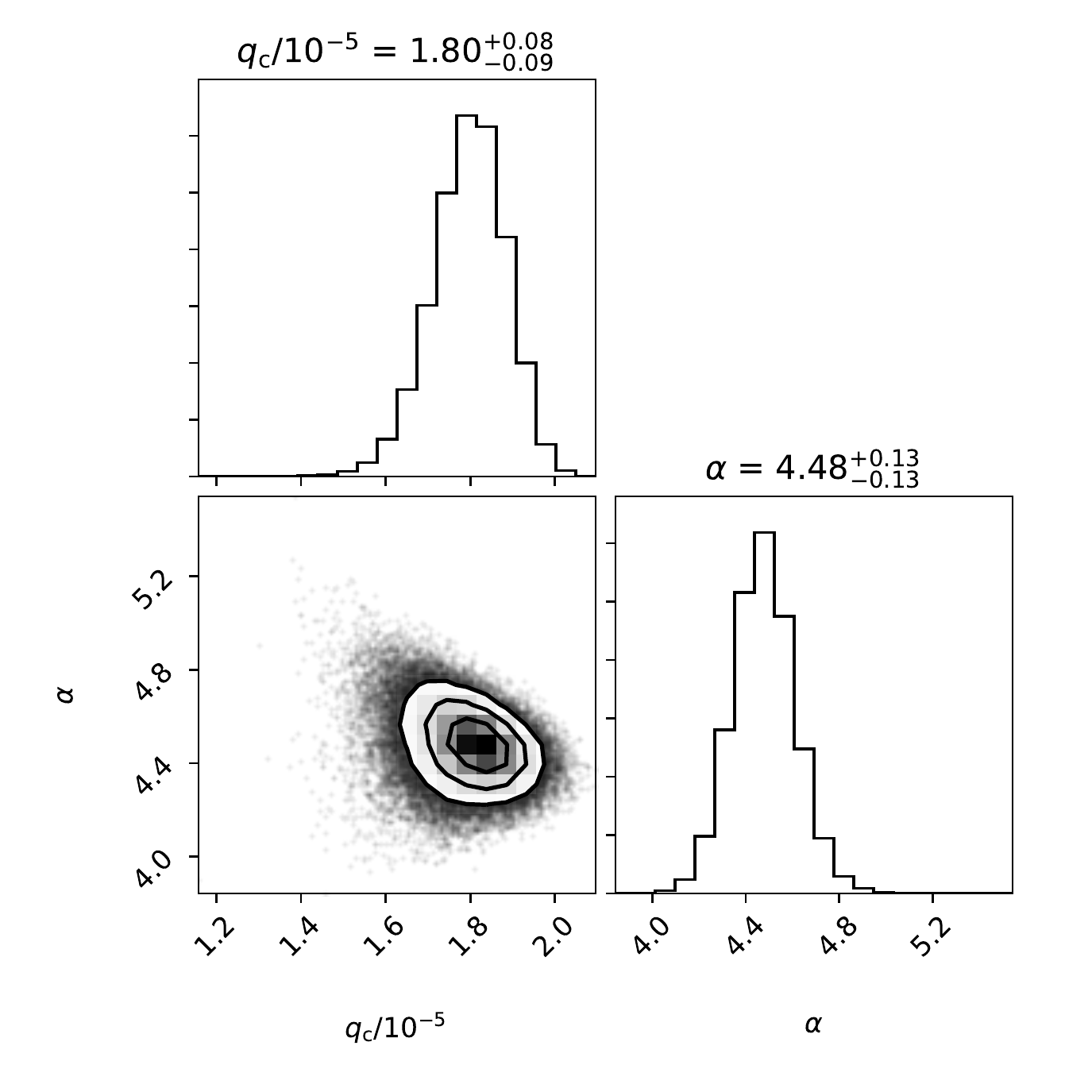}
	\caption{Upper panel: The best fit to the data. The data points are from the Parkes sample, and the error bars are the poisson error. The red dash line denotes the best fitting curve. Low er panel: The  posterior probability distribution of the fitting parameters.}
	\label{fig:5}
\end{figure}

Similarly, the fitting results of samples of CHIME, ASKAP and Parkes are shown in Fig. \ref{fig:3},  Fig. \ref{fig:4} and Fig. \ref{fig:5}, respectively.
We get the parameters of the distribution function $\phi(q)$ as $(q_{\rm{c}}/10^{-5}=1.20^{+0.22}_{-0.52}, \alpha=4.49^{+0.54}_{-0.28})$, $(q_{\rm{c}}/10^{-5}=1.33^{+0.23}_{-0.48}, \alpha=4.34^{+0.63}_{-0.47})$ , $(q_{\rm{c}}/10^{-5}=2.02^{+0.06}_{-0.07}, \alpha=4.82^{+0.14}_{-0.13})$ and $(q_{\rm{c}}/10^{-5}=1.80^{+0.08}_{-0.09}, \alpha=4.48^{+0.13}_{-0.13})$ for the samples of UTMOST, CHIME, ASKAP, and Parkes, respectively.
It can be seen that the samples of ASKAP and Parkes give better constraint to the model parameters.  
However, due to the relatively small size of the samples of UTMOST and CHIME, the constraint to the model parameters are relatively weak.
Nevertheless, one can average the fitting results, because those four observed samples can be regarded as independent experiments. One gets $\bar{q}_{\rm{c}}/10^{-5}=1.59^{+0.08}_{-0.18}$ and $\bar{\alpha}=4.53^{+0.21}_{-0.14}$.

When fitting the data above, we also get the local event rate density of the bursts .
Following Ref. \cite{2018PhRvD..98l3016D}, calibrating with the sample of Parkes, we get the local event rate density with charge larger then ${q_{\min }}$ i.e. 
${n_{0}(> {q_{\min }})} \sim {10^3} {\rm{Gp}}{{\rm{c}}^{ - 3}}{\rm{y}}{{\rm{r}}^{ - 1}} $, where $q_{\rm{min} } \simeq 6 \times 10^{-6}$ is the minimum value of Parkes sample.
Then once knowing $\phi(q)$, we can use equation (\ref{eq-n0}) to calculate the abundance of primordial black holes with charge greater than $q_{\rm{min}}$ i.e. $f(>q_{\rm{min}})$ to account for the event rates of the bursts. We have  $f(>q_{\rm{min}}) \sim 10^{-5}$ for ${n_{0}(> {q_{\min }})}  \sim {10^3} {\rm{Gp}}{{\rm{c}}^{ - 3}}{\rm{y}}{{\rm{r}}^{ - 1}}$.
This means that if the abundance of the primordial black hole populations with charge greater than ${q_{\min }}$ is larger than $10^{-5}$ which is far below the upper limit given by current observations for the abundance of  Earth-mass primordial black holes $f \lesssim 10^{-2}$ \citep{2007A&A...469..387T,2009ApJ...706.1451M,2016PhRvD..94h3504C,2019PhRvD..99h3503N}, then the FRB event rate can be explained in the context of this model. 

\begin{table}[tb]
	\centering
	\caption{The fitting value of the model parameters for different FRB samples, as well as  their average. The errors are all at the 1 $\sigma$ confidence level.}
	\begin{tabular}{ccccc}
		\hline
		\hline
		Samples  &~~~~~~~~~~~~~~~& $q_{\rm{c}}/10^{-5}$ &~~~~~~~~~~~~~~~& $\alpha$ \\
		\hline
		UTMOST &~~~~~~~~~~~~~~~& $1.20^{+0.22}_{-0.52}$ &~~~~~~~~~~~~~~~& $4.49^{+0.54}_{-0.28}$ \\
		CHIME &~~~~~~~~~~~~~~~& $1.33^{+0.23}_{-0.48}$&~~~~~~~~~~~~~~~& $4.34^{+0.63}_{-0.47}$ \\
		ASKAP &~~~~~~~~~~~~~~~& $2.02^{+0.06}_{-0.07}$ &~~~~~~~~~~~~~~~& $4.82^{+0.14}_{-0.13}$\\
		Parkes &~~~~~~~~~~~~~~~& $1.80^{+0.08}_{-0.09}$ &~~~~~~~~~~~~~~~& $4.48^{+0.13}_{-0.13}$ \\
	    \hline
	    Average value &~~~~~~~~~~~~~~~& $1.59^{+0.08}_{-0.18}$ &~~~~~~~~~~~~~~~& $4.53^{+0.21}_{-0.14}$ \\
		\hline
		\hline
	\end{tabular}
	\label{tab:Detection}
\end{table}

\section{Conclusion and Discussion}
In this work, strongly inspired by recent progress on possible evidence for Earth-mass primordial black holes,  we revisit the model of Earth-mass primordial black hole mergers as sources for non-repeating FRBs. Under the null hypothesis that the observed non-repeating FRBs are originated from this model, we analyzed four independent samples to study the model parameters i.e. the typical charge value $q_{\rm{c}}$ and  the power index $\alpha$ of the $\phi(q)$ which describe how the charge was distributed in the primordial black hole population. It turns out that this model can explain the observed data well. Combining the fitting results given by the four samples, we get the typical charge value $\bar{q}_{\rm{c}}/10^{-5}=1.59^{+0.08}_{-0.18}$ and the power index $\bar{\alpha}=4.53^{+0.21}_{-0.14}$.
This implies that the charge distribution of the primordial black holes can be described by a single power function. 
The distribution has a typical charge value, and the number of primordial black holes decreases rapidly with the increase of the charge contained.
The  event rate of the non-repeating FRBs can be explained in the context of this model, if the abundance of the primordial black hole populations with charge greater than $q\sim 10^{-6}$ is larger than $10^{-5}$  which is far below the upper limit given by current observations for the abundance of  Earth-mass primordial black holes.
Furthermore, if the results of OGLE are indeed evidence of the  Earth-mass primordial black holes, then their abundance should be $f \approx 10^{-2}$ \citep{2019PhRvD..99h3503N}.
It means that this model works by requiring only a small portion of the primordial black holes to carry a amount of charge $q \gtrsim 10^{-6}$.

{In principle, primordial black holes could be charged.  But as we know, the universe is full of plasma, so the electrically charged black holes can easily be neutralized by the surrounding plasma. If there is any mechanism that keeps the electric charges of the black hole from being neutralized, the Wald mechanism might work as discussed in Ref. \cite{2018PhRvD..98l3016D}. However, in order to maintain a large electric charges through the Wald mechanism, a sufficiently strong external magnetic field is needed. This requires the primordial black holes to be in a special environment, which seems to be unrealistic.
}

{In contrast, the magnetic charges, as well as topological charges, in black hole can exist stably without being neutralized by the standard local electromagnetic processes.
}\cite{2005PhLB..606..251S} noted that  primordial black holes can sufficiently accrete magnetic monopoles in the early universe and carry magnetic charges, which in principle can be up to $q \sim 10^{-4}$\citep{2018PhRvD..98l3016D}.  Some novel ideas have also been proposed, such as magnetic black hole solution is expected in some kind of modified theory of gravity \citep{2020PhRvD.102j4038C},  topologically induced black hole charges through the effects of quantum matter within general relativity with extra quadratic in curvature terms \citep{2020arXiv200804506K}.
Those charges induced  by the mechanisms mentioned above can survive the long evolutionary
history of the entire universe to keep the charges before merger of the primordial blak holes.

{This raises an interesting question. If the primordial black holes had magnetic charges that are needed to explain the FRBs, would their magnetic fields affect those of galaxies? It's easy to figure out that the average distance between the primordial black holes in the present universe is $R \sim (f \rho_{0}/m_{\rm{PBH}})^{-1/3}$, where $m_{\rm{PBH}}$ is the mass of the primordial black holes. 
Therefore,  the magnetic field generated by those magnetic charges, on the scale of R, is 
\begin{equation}
\begin{aligned}
B \sim & 10^{-24}\left(\frac{q}{10^{-5}}\right)\left(\frac{f}{10^{-5}}\right)^{2 / 3} \\
& \times \left(\frac{\rho_{0}}{10^{-31} \rm{g~ c m^{-3}}}\right)^{2 / 3} \left(\frac{m_{\rm{PBH}}}{10^{28} \rm{g}}\right)^{1/3} \rm{G}~.
\end{aligned}
\end{equation}
It can be seen that the magnetic field strength generated by the magnetic charges in the primordial black holes is too small  to  affect any galaxy's large-scale magnetic field in the present universe.
}

Finally, as pointed out in \cite{2018PhRvD..98l3016D},  the gravitational waves produced
by mergers of Earth-mass primordial black holes will be detected by GHz gravitational wave detectors \citep{2000CQGra..17.2525C,2003CQGra..20.3505B,2006AIPC..813.1280B,2008PhRvD..77b2002N,2009PhRvD..80f4013L,2020arXiv201212189H}. Future simultaneous detection of FRBs and high frequency gravitational waves may bring us
surprises.

\section*{Acknowledgments}
This work is supported by the National Natural Science Foundation of China (grant No. 12047550), the China Postdoctoral Science Foundation (grant No. 2020M671876) and the  Fundamental Research Funds for the Central Universities.

\bibliography{ref}

\end{document}